%% file: main.tex
\documentclass[conference]{IEEEtran}
\IEEEoverridecommandlockouts

\usepackage{cite}
\usepackage{amsmath,amssymb,amsfonts}

\usepackage{graphicx}
\usepackage{textcomp}
\usepackage{tikz}
\usepackage{pgfplots}
\usepackage{xcolor}
\usepackage{soul,color}
\usepackage{makecell}
\usepackage[font=small,skip=2pt]{caption}
\usepackage{adjustbox}
\usepackage{hyperref}
\usepackage{array}
\usepackage{paralist}
\usepackage{booktabs}
\usepackage{siunitx}
\usepackage{tablefootnote}
\usepackage{subfigure}
\usepackage{float}
\usepackage{algorithmic}
\usepackage{algorithm}
\DeclareSIUnit\ms{ms}
\def\BibTeX{{\rm B\kern-.05em{\sc i\kern-.025em b}\kern-.08em
    T\kern-.1667em\lower.7ex\hbox{E}\kern-.125emX}}

\usepackage[acronyms,nomain,xindy,nowarn]{glossaries}
\makeglossaries
\loadglsentries{acronyms.tex}
\setacronymstyle{long-short}
\glsdisablehyper

\usepackage{todonotes}

\usepackage{physics}

\def\BibTeX{{\rm B\kern-.05em{\sc i\kern-.025em b}\kern-.08em
    T\kern-.1667em\lower.7ex\hbox{E}\kern-.125emX}}

\title{PNap: Lifecycle-aware Edge Multi-state sleep for Energy Efficient MEC}

\author{\IEEEauthorblockN{Federico Giarrè, Holger Karl}
\IEEEauthorblockA{\textit{Hasso-Plattner Institute (HPI)}, Digital Engineering Faculty,
University of Potsdam\\
Email: federico.giarre, holger.karl at hpi.de}}

\begin{document}

\maketitle

\begin{abstract}

  \glspl{mec} enables low-latency services by executing applications at the network edge.
  To fulfill low-latency requirements of \emph{mobile} users, 
  providers have to keep multiple edge servers  running at multiple locations, even when, in low-load phases, their capacity  is not needed.
  This significantly increases energy consumption.
  Multi-state sleep mechanisms mitigate this issue by allowing servers to enter progressively deeper sleep states, trading energy savings for longer wake-up delays.   At the same time, service execution depends on non-instantaneous lifecycle operations that cannot be performed while servers are asleep, tightly coupling energy management with service continuity. This paper introduces PowerNap (PNap), a lifecycle-aware  orchestration framework that \emph{jointly} manages server sleep states and service lifecycle states. By leveraging traffic forecasting, PNap jointly minimizes the number of active edge servers and service disruptions. We compare PNap against baselines approaches and a state-of-the-art approach. Results validate PNap, showing how it can reduce energy consumption by up to 14.9\%  with respect to a state-of-the-art solution while matching its service availability results.

\end{abstract}

\glsresetall

\section{Introduction}

\gls{mec} architectures enable latency-critical and highly available services by running services at \glspl{ec} situated close to \glspl{bs}. By using   \glspl{ec} close to users, services can meet stringent \glspl{sla} like end-to-end latency. Achieving these benefits at scale, however, requires deploying many \glspl{ec} across the service area and supporting user mobility via frequent service migrations to maintain proximity between user and service locations \cite{giarre_surfing_2025,giarre_ripple_2026}.

To guarantee service availability at low latency, multiple \glspl{ec} may have to remain active even when idle; else, a mobile user's service might have to be executed on far-away nodes, violating latency requirements. This leads to a significant energy expense \cite{koo_mec_2024,koo_mec_2024-2}. Edge sleeping mechanisms address this issue by transitioning unused \glspl{ec} into low-power states. This has been shown to greatly benefit energy efficiency in \gls{mec} networks \cite{koo_mec_2024,koo_mec_2024-2}.
In particular, multi-state sleep as a paradigm allows multiple sleep depths, trading reduced power consumption for longer wake-up and sleep transition times \cite{gu_energy_2020}.

The \emph{availability} of a service, however, depends on more than just the proximity and power state of \glspl{ec}. To process requests in time, an \gls{ec} must also be running the user's required service. But before a service can be run, \glspl{ec} may need to execute a sequence of time-consuming operations, e.g., deployment and startup. These operations, referred to as \emph{lifecycle}, are a key yet overlooked aspect of service placement in MEC networks \cite{giarre_surfing_2025,giarre_ripple_2026}. Ignoring lifecycle during service placement not only reduces the effectiveness of a placement decision itself but can also cause severe service disruptions. Furthermore, lifecycle operations cannot be performed while an \gls{ec} is in any sleep state, tightly coupling lifecycle with edge power management.

\begin{figure}
    \centering
    \includegraphics[scale=0.9]{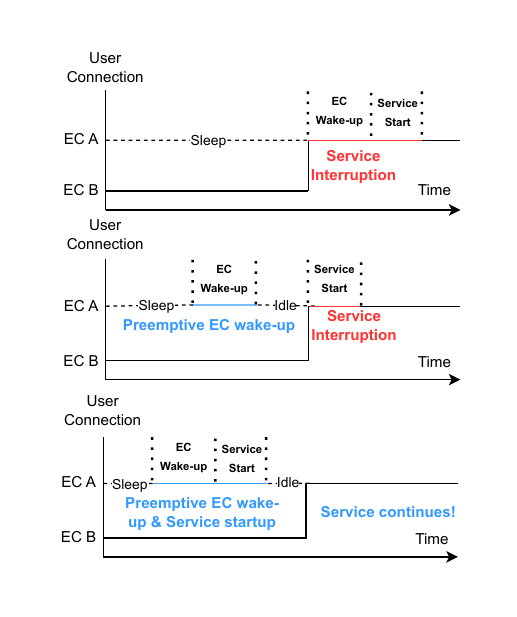}
    \caption{Example of wake-up delay and service lifecycle at user connection.}
    \label{fig:example}
\end{figure}

Taken together, these limitations highlight a key gap in current MEC solutions. To serve users with the requested service and meet latency constraints, \glspl{ec} must be activated ahead of time and lifecycle operations must be initiated proactively, even when no user traffic is currently present. This inevitably increases energy consumption, revealing a fundamental trade-off that cannot be addressed by sleep management or service placement alone. Instead, \gls{ec} sleeping policies and lifecycle-aware service provisioning must be jointly considered.

\autoref{fig:example} illustrates why wake-up/sleep transitions and service lifecycle delays must be jointly considered. If a user is handed over to an inactive \gls{ec} that lacks a running instance of the required service, the combined wake-up and deployment latency can cause service interruptions lasting several seconds. Even when the \glspl{ec} are woken up proactively, the absence of the requested service at handover time still leads to disruption. Only when both the \gls{ec} activation state and service lifecycle are managed proactively can seamless user handover be obtained. Crucially, it is difficult to align preemptive actions and user arrivals, exposing a fundamental trade-off between service continuity and the energy cost of keeping idle \glspl{ec} active while awaiting handovers \cite{koo_mec_2024}.

A key helper to tackle this problem would be  accurate knowledge of future user connectivity, enabling \glspl{ec} to be woken up and services instantiated at the right time and location. But user mobility is inherently uncertain, and future connectivity can only be estimated, often with significant errors. This uncertainty makes it necessary to design approaches that are not only predictive but also robust to imperfect forecasts. Importantly, this work addresses individual service deployment onto \glspl{ec}. 

In this paper, we address these challenges as  follows:
\begin{itemize}
\item We  model the  combination of \gls{ec} multi-state sleep behavior and service lifecycle.
\item We discuss a novel request-handling model for \glspl{ec}, taking into account different types of requests and their management overhead.
\item We formulate a novel multi-objective problem that explicitly balances \gls{ec} energy consumption against \gls{sla} violations.
\item We use a \gls{stgcn} to estimate future user connectivity and use its output to guide a lifecycle and sleep-aware heuristic for service placement and power control.
\end{itemize}

\section{Related Work}
\label{sec:related-work}

Several studies address \gls{mec} server sleep. None of these, however,  addressed the usage of the multi-state sleep paradigm in the context of \gls{mec}, nor how it ties with lifecycle operations, both crucial to correctly serve users.

For example, Koo and Park \cite{koo_mec_2024,koo_mec_2024-2} propose a traffic-prediction-based framework that dynamically adjusts the number of active MEC servers, predicting  traffic using a \gls{stgcn}, achieving significant energy savings while meeting delay requirements. While effective, the authors do not model transition delays between \gls{ec} state, assuming slotted time. Levis \emph{et al.} \cite{levis_sleepy-rapp_2025} introduce an  orchestration framework, \emph{SLEEPY-rApp} that selectively deactivates servers and reroutes tasks to active neighbors to reduce energy consumption under latency constraints.
The authors consider reactivation time of \glspl{ec}, but they ignore the time taken  to transit to the sleep state.
Javed \emph{et al.} \cite{javed_energy-efficient_2025} formulate MEC server activation as a minimum set-cover problem to identify the smallest subset of active servers required to satisfy latency constraints.
It is closely related to coverage-based formulations yet largely static and does not incorporate user mobility or finite-horizon sleep decisions.
Hou \emph{et al.} \cite{hou_efficient_2024,hou_intelligent_2024} investigate server sleep mechanisms in vehicular networks with the dual objective of improving energy efficiency and resource utilization. Their solutions rely on classical clustering techniques \cite{hou_efficient_2024} and deep reinforcement learning approaches \cite{hou_intelligent_2024}; however, transition times, multi-state sleep models and lifecycle effects are not considered.
Park and Lim \cite{park_bio-inspired_2023} address energy-efficiency optimization under service delay constraints using a bio-inspired genetic algorithm that enables edge clouds to coordinate sleep decisions with neighboring nodes. In their model, state transitions are assumed to be instantaneous, and neither multi-state sleep mechanisms nor lifecycle dynamics are examined.
Wang \emph{et al.} \cite{wang_cooperative_2019} propose a Lyapunov-based approach to decide which \glspl{ec} to keep active to maximize energy efficiency. In the proposed model, transition delays are not discussed, nor are  multi-state sleep or lifecycle aspects.

Lifecycle-aware service provisioning has, to our knowledge, only been addressed  by us \cite{giarre_surfing_2025,giarre_ripple_2026}. Our previous study showed the importance of considering lifecycle management during service deployment to not interrupt  services during user mobility. It did not, however, take into account energy efficiency and \gls{ec} sleep modes. 

\autoref{tab:rw} proposed a summary of the work discussed in this section. Prior work demonstrates the effectiveness of MEC server sleep strategies for improving energy efficiency but generally neglects the non-negligible server state transition times and lifecycle operation. This work addresses these limitations by modeling MEC server multi-sleep as a time-dependent coverage problem tied with the problem of service provisioning under realistic lifecycle requirements.

\begin{table*}[ht!]
    \centering
    \caption{Summary of the related work section}
    \label{tab:rw}
    \begin{tabular}{lccccc}
    \toprule
        Study & Objective & Solver & Transition delays & Multi-state sleep & Service lifecycle \\
    \midrule
       This study &Energy efficiency, service availability&\gls{stgcn}+Heuristic&\checkmark&\checkmark&\checkmark\\
       Koo and Park \cite{koo_mec_2024} &Energy efficiency&\gls{stgcn}+DQN&x&x&x\\
       Koo and Park \cite{koo_mec_2024-2} &Energy efficiency, delay&STGCN+Heuristic&x&x&x\\
       Levis \emph{et al.} \cite{levis_sleepy-rapp_2025}  &Energy efficiency, delay &Heuristic&\checkmark&x&x\\
       Javed \emph{et al.} \cite{javed_energy-efficient_2025}  &Energy efficiency, service availability&Heuristic&\checkmark&x&x\\
       Hou \emph{et al.} \cite{hou_efficient_2024}  &Energy efficiency, resource utility &Heuristic&x&x&x\\
       Hou \emph{et al.} \cite{hou_intelligent_2024}  &Energy efficiency&DRL&x&x&x\\
       Park and Lim \cite{park_bio-inspired_2023}  &Energy efficiency, delay&Genetic algorithm&x&x&x\\
       Wang \emph{et al.} \cite{wang_cooperative_2019}  &Energy efficiency&Lyapunov&x&x&x\\
       Giarrè and Karl  \cite{giarre_surfing_2025,giarre_ripple_2026} &Service availability&LSTM+RF+Heuristic&\checkmark&x&\checkmark\\
   
    \bottomrule
    \end{tabular}
    
\end{table*}

\section{System Model}\label{sec:system}
We consider a set  $\mathcal{U}$ of users moving in a network composed of a set  $\mathcal{E}$ of \glspl{ec}. Each \gls{ec} covers a section of the network, without overlapping with others' coverage areas; users in a coverage area automatically connect to that \gls{ec}.
Each user requests a service $s\in\mathcal{S}$,
which can be deployed at any \gls{ec}. Users may connect to any instance of the service throughout the network, even if hosted by a different \gls{ec} distant from the user. Each user has a load $\lambda_u$ of requests per second that have to be processed by its service within a deadline. \glspl{ec} have a limited array of computational resources available, i.e. $R_\mathrm{EC}$, with each service consuming  $R_s$ of them; we consider a vector of resource types at least comprising CPU and memory. 

\subsection{Service Lifeycle}

\begin{figure}[t]
    \centering
    \includegraphics[scale=0.8]{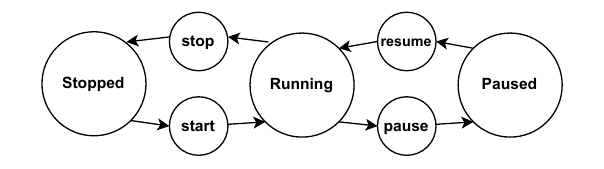}
    \caption{Lifecycle of a service\cite{giarre_surfing_2025}}\label{fsm}
  \end{figure}

\autoref{fsm} shows a \gls{fsm} that generalizes the mechanism of starting, stopping and pausing a service. In this work, we consider only a subset of the states considered in our previous work \cite{giarre_surfing_2025,giarre_ripple_2026}, which models the entirety of states through which a service has to pass before becoming operational. These states are common in the vast majority of virtualized software, like \glspl{vm}\footnote{\url{https://docs.openstack.org/nova/latest/reference/vm-states.html}} and containers \cite{stahlbock_optimization_2022} but can be easily extended to any software.

We assume here that each service at \glspl{ec} is always \emph{at least} in the stopped state. We argue that it is realistic for \glspl{ec} to have deployments ready for services supported by the network provider. 
Transitioning between two states $\psi$ and $\psi'$ requires a time $t_{\psi,\psi'}$ and, during this time, users cannot be served by the service. Additionally, we consider that service state transitions cannot be interrupted.
Different service states require different  resources  (\autoref{tab:usage}), affecting which and how many services can be deployed in which state at a resource-constrained \gls{ec}.

\begin{table}[]
    \centering
    \caption{Resources usage at each state}
    \begin{tabular}{lcc}
        \toprule
        State &   CPU & Memory  \\
        \midrule
        Stopped &  x & x \\
        Paused   & x & $\checkmark$\\
        Running & $\checkmark$ & $\checkmark$ \\
        
        \bottomrule
    \end{tabular}
    
    \label{tab:usage}
  \end{table}

\subsection{End-to-end latency computation}\label{latency}
A user's \gls{e2e} latency $t_u = t_\text{wireless} + t_f + t_s$ is the sum of:
\begin{inparaenum}[(i)]
    \item wireless link latency $t_{\text{wireless}}$;
    \item network forwarding $t_{f}$;
    \item service time $t_s$.
\end{inparaenum} 
This latency $t_u$ should be smaller than a  service's \gls{e2e} latency limit $t_{\max{}}$.

\subsubsection{Wireless delay}

We model the wireless delay $t_{\text{wireless}}$ as a constant \cite{coll-perales_end--end_2022}. Although simplified, this  is a reasonable model since the wireless access link does not meaningfully affect the rest of our model and is not influenced by our proposed approach. In addition, for the considered class of low-latency \& highly dependable  services,
 request packets are typically small, allowing transmission and propagation delays to be neglected \cite{firouzi_delay-sensitive_2024}.

\subsubsection{Network forwarding}


We account  for the queueing delay $t_q$ on links and the packet processing time at the \glspl{ec}, denoted by $t_p$ \cite{firouzi_delay-sensitive_2024}. Packet processing time $t_p$ is a constant added at every forwarding hop:  the time to receive and send request packets \cite{coll-perales_end--end_2022}. Queueing time  $t_q$ on link $l$ is modeled as an M/D/1 queue with service rate $\mu_l$ and  arrival rate $\lambda_l$, given by the aggregate traffic generated by all users whose packets traverse that link. In summary: $ t_f~=~\sum_{l\in\mathcal{L}} \left(\frac{2-\rho}{2\mu_l(1-\rho)} + t_p\right) \quad \text{with} \quad \rho=\frac{\lambda_l}{\mu_l}$ and with $\mathcal{L}$ the  links through which  requests go from the user to the service. This penalizes assigning users to distant service locations.



\subsubsection{Service time at \glspl{ec}}
Each \gls{ec}  has $c$ processing cores available to handle incoming requests. In addition to user-generated service requests, \glspl{ec} also receive control requests, such as lifecycle commands from the network orchestrator to change service states or the \gls{ec}'s activity state. These control requests fundamentally differ from user requests as they do not follow a non-markovian arrival process. As a result, conventional queueing models such as (variants of) M/M/$c$ 
are not directly applicable. 
We use separate queues for different types of requests. Service across queues is scheduled using a preemptive priority discipline, with requests strictly ordered by priority, while requests in the same queue first-in first-out (FIFO) policy. Naturally, as higher-priority requests are being processed, the cores handling them are temporarily unavailable to lower-priority requests, potentially increasing their service times. Such preemptive scheduling between FIFO queues serves well to capture the abstraction level of lifecycle management. More refined policies are of course possible but left for further study.

We consider $N$ priority queues $i=1, \dots, N$. There are   $q_i$ requests of class $i$ in the system (being served or queued), with  $k_i$ cores running requests of class $i$, bounded by the total number of cores: $\sum_{n\in N}~k_n~\leq~c$.  The system state is hence $(q_1,\dots,q_N,k_1,\dots,k_N)$. 
The total number of cores in use is bounded by the total number of cores $\sum_{n\in N}~k_n~\leq~c$. An arriving request is serviced immediately if there is an idle core or if a lower-priority request can be preempted; else it has to wait. When preempted, a request is put in front of its queue, waiting to be served for the remaining service time. Once a request finishes processing, a waiting request with the same priority starts processing immediately or, if none exists, a lower-priority request does. When a request finishes processing, no higher-priority request can be waiting (it would have preempted running the request), so only a same-priority or lower may start processing. If no request is waiting, cores idle.

\subsection{Power consumption}
The power consumption of an \gls{ec} is influenced by several factors. In this work, we focus on two in particular: the processing load when the \gls{ec} is active and the power behavior associated with sleep states and state transitions.

\subsubsection{Processing Load}
We denote by $P_\text{idle}$ the power consumption of an active \gls{ec} when no user requests are being processed and by $P_\text{peak}$ its maximum power consumption. The instantaneous power drawn by \gls{ec} $e$ increases linearly with the number of cores  $c_e$ currently in use, offset by the idle power~\cite{hou_efficient_2024}:
\begin{equation}\label{eq:power}
P_e = P_\text{idle} + c_e \left(P_\text{peak} - P_\text{idle}\right),
\end{equation}

\subsubsection{\gls{ec} State and Transitions}
The previous model captures power consumption when \glspl{ec} are active, either under load or idle. When an \gls{ec} enters a sleep state, however, its power consumption decreases substantially. Following \cite{gu_energy_2020}, we consider a set of sleep states $1 \dots \Phi$ ordered by increasing sleep depth, each characterized by a constant power consumption $P_{\phi}$, with $P_\text{peak} \geq P_\text{idle} \geq P_{1} \geq \dots \geq
P_\Phi 
$.

Transitions between the active state and sleep states are not instantaneous and incur both time and energy overheads. We denote by $t_{\text{active},\phi}$ the time required to transition from the active state to sleep state $\phi$, with deeper sleep states requiring longer transition times, i.e., $t_{active,1} \leq \dots \leq t_{\text{active},\Phi}$.
Symmetrically, waking up from deeper sleep states takes longer. During these transitions, power consumption is higher than in the corresponding sleep state, denoted by $P_{1,\text{active}} \geq \dots \geq P_{\Phi,\text{active}}$.

We consider, according to the current ACPI standard for server sleep modes\footnote{\url{https://uefi.org/htmlspecs/ACPI_Spec_6_4_html/16_Waking_and_Sleeping/sleeping-states.html}}, that direct transitions between different sleep states are not allowed; \glspl{ec} can only transition between the active state and a sleep state and vice-versa. Additionally, once a state transition is started we consider that is not possible to abort it. Finally, we assume that the wake-up process includes resuming all services to their state prior to entering the sleep state.

\section{Problem Formulation}
\label{sec:problem-formulation}

With the intent of minimizing energy consumption throughout the infrastructure while complying with the services' \gls{sla}, we formulate the following optimization problem. Let $P_e^t$ the power consumed at \gls{ec} $e$ at time $t$, and $v_u^{r,t}$ a binary variable equal to 1 if request $r$ of user $u$ either surpasses the latency limit or is dropped at time $t$, then we model the following objective function:
\begin{equation}
  \label{eq:energy_consumption}
    \min \lim_{T\rightarrow\infty} \quad \frac{1}{T}\sum_{t}^{T}  \left[\omega_p\sum_e^{\mathcal{E}} P_e^t + \omega_v\sum_u^U\sum_r^{R_u} v_u^{r,t} \right]
\end{equation}
With $\omega_p$ and $\omega_v$ weights for the individual objectives.

Let $z^t_{e,s,\psi}$ be the binary variable set to 1 if service $s$ at \gls{ec} $e$ is in state $\psi$ at time $t$, and $c^t_{u,e}$ the binary variable set to 1 if user $u$ is connected to \gls{ec} $e$ at time $t$. We use $l_{e,s,\psi}^t=1$ if service $s$ at \gls{ec} $e$ is currently transitioning to state $\psi$. Let $st^t_{e,\phi}$ be the binary variable indicating whether \gls{ec} $e$ is in state $\phi$ at time $t$ and $q_{e,\phi}^t$ the binary variable 1 if \gls{ec} $e$ is transitioning to state $\phi$ at time $t$.
Any solution to this problem must comply with these constraints at any time $t\in\{0...T\}$:

\begin{subequations}
    
    \begin{equation}\label{x1}
        \sum_e^\mathcal{E} c_{u,e}^t = 1 \quad \forall u \in \mathcal{U},
    \end{equation}
    \begin{equation}\label{x4}
        c_{u,e}^t \leq z_{e,s_u,\text{running}}^t \cdot st^t_{e,\text{active}} \quad \forall u\in \mathcal{U}, e \in \mathcal{E} \footnote{This constraint, like \eqref{l1} and \eqref{l2}, is expressed in quadratic form but can be readily decomposed into multiple constraints for linearization.} 
    \end{equation}
    
    \begin{equation}\label{z1}
        \sum_{\psi\in\Psi} z_{e,s,\psi}^t = 1 \quad \forall e \in E, s \in \mathcal{S},
    \end{equation}
    \begin{equation}\label{r1}
        \sum_{v\in\mathcal{V}} R^t_{e,s} \leq R_{tot} \quad \forall e \in \mathcal{E},
    \end{equation}
    \begin{equation}\label{r2}
        \rho_e^t < 1 \quad \forall e \in \mathcal{E}
    \end{equation}
    \begin{equation}\label{l1}
        l_{e,s,\psi'}^{t+\delta} \geq z_{e,s,\psi}^{t-1} \cdot z_{e,s,\psi'}^t \; \forall e \in \mathcal{E}, s \in \mathcal{S}, \psi \neq \psi', \delta \in \{0... t_{\psi,\psi'}\},
      \end{equation}
      \noindent with $t_{\psi,\psi'}$ the transition time between the two states $\psi$ and $\psi'$ in the lifecycle \gls{fsm}, cumulative in case $\psi$ and $\psi'$ are not directly connected. 
    \begin{equation}\label{l2}
        q_{e,\phi'}^{t+\delta} \geq st_{e,\phi}^{t-1}\cdot st_{e,\phi'}^{t} \quad \forall e \in \mathcal{E}, \phi \neq \phi', \delta \in \{0 \dots t_{\phi,\phi'}\}
    \end{equation}
    with $t_{\phi,\phi'}$ the transition time between two \glspl{ec} states.
    
    Constraints \eqref{l1} and \eqref{l2} act as locking mechanisms: when a change occurs in the service state or in the activity state, respectively, the corresponding variables are iteratively set to 1 over the transition interval. This forces the solver to maintain the new state decision for the transition duration, as enforced by constraints \eqref{l3} and \eqref{l4}.
      \begin{equation}\label{l3}
        z_{e,s,\psi}^t \geq l_{e,s,\psi}^t \quad \forall e \in \mathcal{E}, s \in \mathcal{S}, \psi \in \Psi,
    \end{equation}
    \begin{equation}\label{l4}
        st^t_{e,\phi} \geq q^{t}_{e,\phi} \quad \forall e \in \mathcal{E}, \phi \in \Phi 
    \end{equation}

    \begin{equation}\label{st1}
        \begin{split}
        st^t_{e,\phi}=1 \implies z^t_{e,s,\psi} \cdot z^{t-1}_{e,s,\psi} = 1  & \\  \forall e \in \mathcal{E}, s \in \mathcal{S}, \psi \in \Psi, \phi \in \Phi\setminus\text{active}
        \end{split}
    \end{equation}

    \begin{equation} \label{v1}
         (c_{u,e}^t \wedge (l_{e,s_u,\text{running}}^t\vee q_{e,\text{active}}^t)) \implies v_u^r = 1\; \forall u\in\mathcal{U}, \forall r \in R_u
    \end{equation}
    \begin{equation}\label{v2}
        t_{u} \leq t_{\max{}} + M\cdot v_u^{r,t} \quad \forall u \in U, \forall r \in R_u,
    \end{equation}
    \noindent where $M$ is a Big-M constant. 
\end{subequations}

\begin{figure*}[htbp]
    \centering
    \includegraphics[scale=0.5]{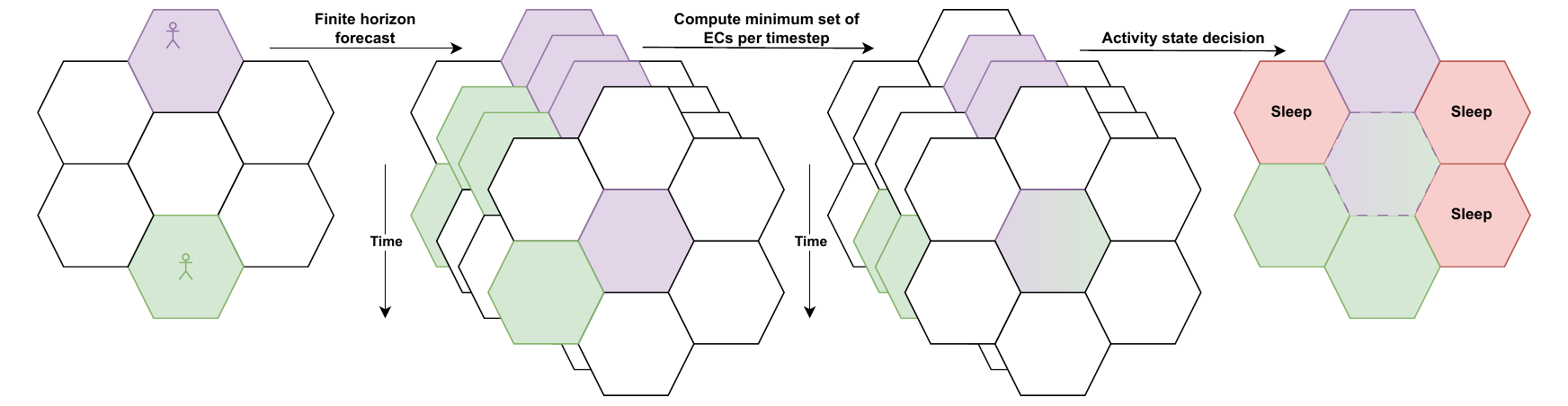}
    \caption{Example of PNap time-evolving coverage approach.}
    \label{fig:pnap}
  \end{figure*}

Constraints \eqref{x1}--\eqref{z1} force users to connect to only one \glspl{ec} where their service is running. Additionally, each service can be in only one state in each \gls{ec}. Constraints \eqref{r1} and \eqref{r2} express resource consumption, where instantiated services cannot allocate more resources than available and the total load of user requests cannot exceed the available service rate. Constraints \eqref{l1} and \eqref{l2} record the start of the transition between two states and the time it takes to complete, while constraints \eqref{l3} and \eqref{l4} ensure that for the duration of the transition neither services nor \glspl{ec} can change their goal state or abort the transition. Similarly, constraint \eqref{st1} indicates that when an \gls{ec} is in a sleep state, no service can change state. Finally, constraints \eqref{v1} and \eqref{v2} set the SLA violation binary variable when the user is trying to connect to a service still transitioning to the running state to an \gls{ec} still transitioning to the active state or when the delay threshold is surpassed.

\section{Proposed Approach}
\label{sec:proposed-approach}

To jointly minimize \gls{sla} violations and energy consumption, MEC orchestration must explicitly coordinate two tightly coupled proactive decisions:
(i) when and where to start or stop service instances, accounting for non-negligible lifecycle delays \cite{giarre_surfing_2025}; and
(ii) when and which \glspl{ec} to wake up or put into multi-level sleep states to serve users, accounting for transition overheads \cite{gu_energy_2020}.
The first decision is closely related to the service provisioning problem, which is NP-hard \cite{attaoui_vnf_2023} even when lifecycle management is not considered, while the second is related to the set coverage problem, which is NP-hard \cite{caprara_algorithms_2000} even without accounting for multi-state sleep modes and state transitions. Both decisions need to be taken in advance to be effective, due to the transition time constraints, which requires anticipating future user connectivity. But exact future knowledge is rarely available, and user-level forecasting and end-to-end learning-based orchestration introduce significant computational and operational overheads \cite{koo_mec_2024}. As a result, optimal solutions and fully data-driven approaches for this problem cannot scale to large networks. We therefore adopt a hybrid approach that combines lightweight learning for forecasting with scalable heuristics for decision making.

To estimate future connectivity, we rely on an \gls{stgcn}. By jointly modeling the spatial structure of the MEC network and the temporal evolution of connectivity, \glspl{stgcn} effectively capture how mobility-induced demand propagates across neighboring \glspl{ec}. Importantly, this model operates at the node level, representing \glspl{ec} in this work, rather than tracking individual users, which significantly improves scalability while retaining sufficient accuracy for proactive orchestration\cite{koo_mec_2024}.

The connectivity estimation is then used to solve a time-evolving coverage problem over a finite prediction horizon. For each forecasted time step, we compute the set of \glspl{ec} required to serve the predicted demand while satisfying latency constraints. The proposed coverage algorithms are detailed in Algorithms~\autoref{algo} and~\autoref{algo2} (following page).   Unlike conventional approaches that solve coverage once for the current network load, our method aggregates coverage decisions across the entire horizon. This aggregation acts as a conservative mechanism: by focusing on \glspl{ec} that are (predicted to be) not required for prolonged periods, the approach mitigates the impact of forecasting errors and reduces the risk of premature shutdowns or delayed activations. Additionally, the clustering behavior can be adjusted by modifying line 5 of Algorithm~\autoref{algo2}. This, as discussed later in Section \ref{sec:numerical}, allows the clustering process to be made more or less conservative trading off energy consumption for service availability and vice-versa to  better match specific use-case requirements. From this perspective, PNap can be interpreted as a configurable framework rather than a fixed, stand-alone solution, enabling adaptation to different context-dependent objectives.

Finally, \glspl{ec} currently in use are kept active, avoiding disrupting users' service while other \glspl{ec} and services are prepared.
\autoref{fig:pnap} shows an example of the proposed approach.

This horizon aggregation directly informs power and service management decisions. \glspl{ec} that are expected to be needed in the near future are kept active or transitioned into shallow sleep states, avoiding costly wake-up delays. Conversely, \glspl{ec} that never appear in the aggregated coverage sets can be safely placed into deeper sleep states to maximize energy savings. Once the set of future active \glspl{ec} and their associated services is identified, service lifecycle operations and user association are triggered proactively, ensuring that services are available at the right place and time when users arrive. 

Correct management of lifecycle operations is essential to support user mobility. Building on the processing model introduced in Section \ref{latency}, we assign higher priority to lifecycle operations than to user requests, avoiding preemptions that could delay the availability of services and cause significant service disruptions, as later shown in Section \ref{sec:numerical}.

The proposed coverage algorithm has a computational complexity, dominated by the clustering algorithm, of $\mathcal{O}(\mathcal{|S|} \cdot |\mathcal{E}|^2)$, where $\mathcal{E}$ is the set of \glspl{ec} and $\mathcal{S}$ the set of services available. As a result, the approach scales well in practice while avoiding the prohibitive complexity of optimal solutions or ML approaches with user-level forecasting.

\begin{algorithm}
 \caption{Coverage algorithm}
 \begin{algorithmic}[1]\label{algo}
 \renewcommand{\algorithmicrequire}{\textbf{Input:}}
 \renewcommand{\algorithmicensure}{\textbf{Output:}}
 \REQUIRE Current placement, current server state, max feasible distance
 \ENSURE Coverage set
 
 \textit{Get the current network load, it has shape $(|\mathcal{E}|,|\mathcal{E}|,|\mathcal{S|})$: the request load is stored for each EC,  from which EC this load is rerouted, and  which service is used.}\\
 \STATE load = getLoad() 
 \textit{Compute, for each EC, the set of nodes reachable within the delay limit}
 \STATE rs = computeReachabilitySet()
 \STATE change = True
 \WHILE{change}
    
    \STATE \textit{Preparation: sort the list of EC by load of users in their coverage area, heading with EC that are not in the active state}\\
    \STATE nodes = nodes.sort()
    \FOR{$n$ in nodes}
        \IF{sum(load[$n$]) $>$ 0}
             \STATE \textit{Try offloading the current load to another EC in the reachability set}
            \STATE change, newload = offload($n$,load,rs)
            \IF{change}
                \STATE load = newload
                \STATE break
            \ENDIF
        \ENDIF
    \ENDFOR
 \ENDWHILE
 \STATE coverageSet = set()
 \FOR{$n$ in nodes}
    \IF{sum(load[$n$])$>$0}
        \STATE coverageSet.add($n$)
    \ENDIF
 \ENDFOR
 \RETURN coverageSet 
 \end{algorithmic} 
 \end{algorithm}

 \begin{algorithm}
 \caption{Offloading algorithm}
 \begin{algorithmic}[1]\label{algo2}
 \renewcommand{\algorithmicrequire}{\textbf{Input:}}
 \renewcommand{\algorithmicensure}{\textbf{Output:}}
 \REQUIRE EC to offload $n$, its reachability set rs, current network load
 \ENSURE success flag, updated network load
 \\
 
 \STATE newload = load.copy()
 \FOR{node $q$ in newload[$n$] \textbf{for} service $s$ in load[$n$][$q$]}
    \IF{load[$n$][$q$][$s$]$>$0}
        \FOR{node $w$ in rs[$n$] \textbf{if} $w$ is in rs[$q$]}
            \IF{$w$ can support the load of $s$}
                \STATE newload[$w$][$q$][$s$] += newload[$n$][$q$][$s$]
                \STATE newload[$n$][$q$][$s$] = 0
                \STATE \textbf{break}
            \ENDIF
        \ENDFOR
    \ENDIF

 \ENDFOR
 \STATE flag = False
 \IF{sum(newload[$n$])==0}
    \STATE flag = True
 \ENDIF
 \RETURN flag, newload
 \end{algorithmic} 
 \end{algorithm}

\section{Numerical Evaluation}\label{sec:numerical}

\subsection{Scenario}
\label{sec:scenario}

An event-driven simulator is implemented for the described model, with 25 \glspl{ec} arranged on a grid \cite{koo_mec_2024}. The network serves 100 mobile users following the Gaussian-Markov mobility model. Each user uniformly requests one of 8 available services, with per-user request arrivals modeled as a Poisson process at 100 requests per second. Each \gls{ec} has resources  $R_\text{EC}=(5,6)$, representing CPU and memory resources, while each service requires $R_s=(1,1)$ to meet its \gls{sla}, regardless of the number of connected users. Importantly, resources in this scenario are specifically constrained to highlight the trade-off between energy consumption and SLA violation.

\subsubsection{Activity and sleep states}
Following \cite{gu_energy_2020}, we consider three sleep states from the ACPI standard:
 S1, S3, and S4. States S2 and S5 are excluded as S2 offers only minimal  savings over S1, and S5 may degrade hardware over time. \autoref{energy} summarizes power consumption and transition delays for all relevant states and transitions.

\begin{table}[]
    \centering
    \caption{Power consumption  and state-transition delays for \gls{ec} in different states \cite{gu_energy_2020}}
    \label{energy}
    \begin{tabular}{cc|ccc}
        \toprule
        State & Power (W) & State & Power (W) & Delay (s) \\
        \midrule
        Peak & 243 & Active $\rightarrow$ S1 & 140 & 2\\
        Idle & 150 & Active $\rightarrow$ S3 & 100 & 4\\
        S1 & 133 & Active $\rightarrow$ S4 & 60 & 9\\
        S3 & 97 & S1 $\rightarrow$ Active & 144 & 2\\
        S4 & 49 & S3 $\rightarrow$ Active & 128 & 10\\
        - & - & S4 $\rightarrow$ Active & 95 & 48\\

        \bottomrule
    \end{tabular}
  \end{table}
  
\subsubsection{Service Lifecycle}
We consider the service states Stopped, Running, and Paused.  Services are deployed through lightweight virtualization, in particular containers. Fu \emph{et al.} \cite{fu_fast_nodate} provide experimental results for the transition times needed for these lifecycle operations. According to their values, we consider the average time to start a container $\hat{t}_{\text{start}}=510$ ms, $\hat{t}_{\text{pause}} = \hat{t}_{\text{resume}} = 96$ ms. Since no result for the transition from the running state to the stopped is given, we consider $\hat{t}_{\text{stop}} = \hat{t}_{\text{start}}$ for a graceful stop of the service.

\subsubsection{Processing at \glspl{ec}}

We implement the queueing model described in Section \ref{latency} as a continuous time markov chain (CTMC). A high-priority queue holds operator-injected service lifecycle requests;  the low-priority one has a Poissonian arrival process for user-generated service requests. The system state is described by the tuple $(h_\text{wait},l_\text{wait},h_\text{run}, l_\text{run})$, where $h$ and $l$ represent the number of high- and low-priority requests currently either waiting or running  at an EC, respectively.

\subsubsection{\gls{stgcn}}

The \gls{stgcn} used to perform connectivity forecasting is trained with the DCRNN \cite{li_diffusion_2018} algorithm using the Torch Spatiotemporal \cite{cini_torch_2022} library, with default training hyperparameters.

\subsection{Comparison Cases}
As discussed in Section~\ref{sec:related-work}, there is no approach in literature that jointly address proactive orchestration of multi-state sleep of \glspl{ec} and service lifecycle. Hence, we need to compare PNap againts plausible approaches, specifically:

\begin{itemize}
\item \textit{Ideal}: \gls{ec} states, user embedding, and service states are determined by an \gls{ilp} solver. This approach doesn not consider constraints \eqref{l1} and \eqref{l2} of the problem formulation and (unrealistically) sets transition delays for both sleep and lifecycle states to 0 s, but with \glspl{ec} power consumption still accounted for. This unrealistic set of assumptions is similar to what is often assumed in prior  work. It hence only serves as an unrealistic performance upper bound. 
    \item \textit{Reactive}: Similar to Ideal it does not consider constraints \eqref{l1} and \eqref{l2} though, conversely from Ideal, transition times for sleep and lifecycle states are still enforced during the simulation. 
    In practice, an \gls{ilp} solver is used to find the best \glspl{ec} and service states as in the Ideal case, but the assumptions about instantaneous transition no longer hold. This baseline illustrates how neglecting these time-consuming processes can degrade system performance in realistic scenarios.
    \item \textit{SLEEPY-like}\cite{levis_sleepy-rapp_2025}: a state-of-the-art heuristic approach to determine a set of \glspl{ec} that can be put to sleep to save energy. SLEEPY addresses the coverage problem in a manner closely aligned with PNap, which makes it a relevant baseline for comparison.  Unlike PNap, however, it neither accounts for multi-state sleep mechanisms nor incorporates service lifecycle operations, relying solely on the instantaneous network state for decision-making. Furthermore, as the original study does not report \gls{ec} sleep state specifications, we assign S2 as its sleeping state to strike a balance between transition time and energy savings.
\end{itemize}

\subsection{Metrics and parameter}
\label{sec:metrics}

All approaches are compared on  \emph{energy} used by \glspl{ec} and \emph{service availability} as metrics, varying the \emph{allowed \gls{e2e} latency} as independent parameter to show how the trade-off between energy consumption and service availability evolves as  the problem's latency requirements are  more or less strict.

\subsection{Results}
\begin{figure}
    \centering
    \includegraphics[scale=0.5]{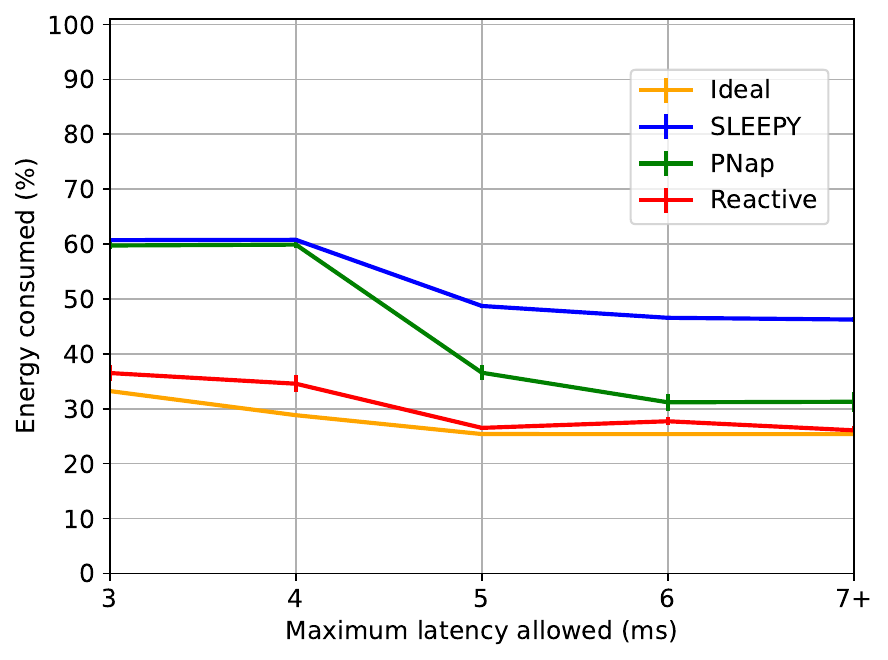}
    \caption{Average percentage of energy consumed over the peak possible consumption with respect to an increasing latency allowed. Standard deviation shown as error bars.}
    \label{fig:energy}
  \end{figure}

\begin{figure}
    \centering
    \includegraphics[scale=0.5]{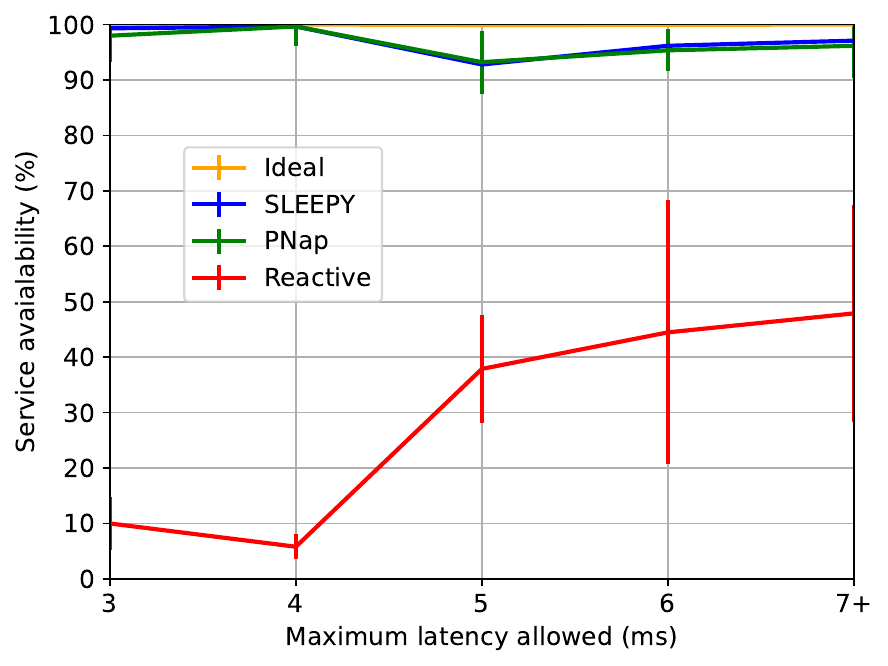}
    \caption{Average user-perceived service availability with respect to an increasing latency allowed. Standard deviation shown as error bars.}
    \label{fig:sa}
  \end{figure}

\begin{figure}
    \centering
    \includegraphics[scale=0.5]{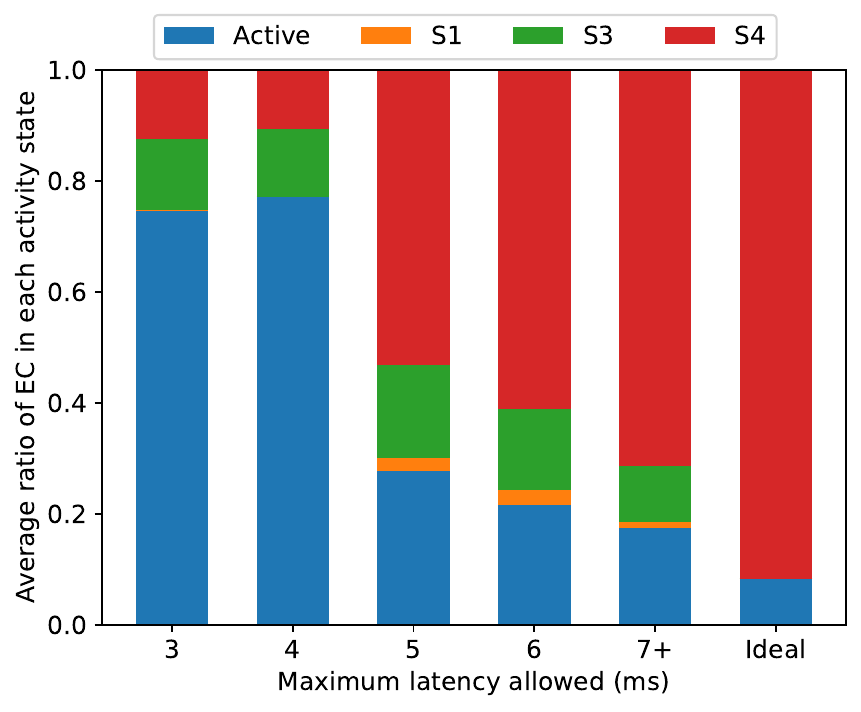}
    \caption{Average ratio of ECs in a particular activity state over simulation time with respect to increased allowed latency.}
    \label{fig:state}
\end{figure}
\begin{figure}
    \centering
    \includegraphics[scale=0.5]{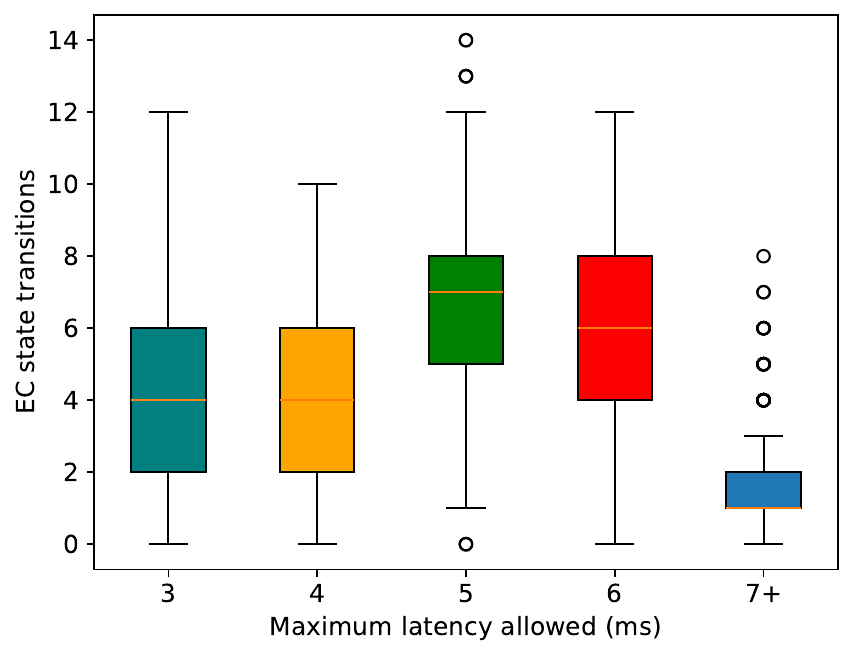}
    \caption{Number of EC state transition over simulation time with respect to increased allowed latency.}
    \label{fig:transition}
\end{figure}

\begin{figure}
  \centering
  \includegraphics[scale=0.5]{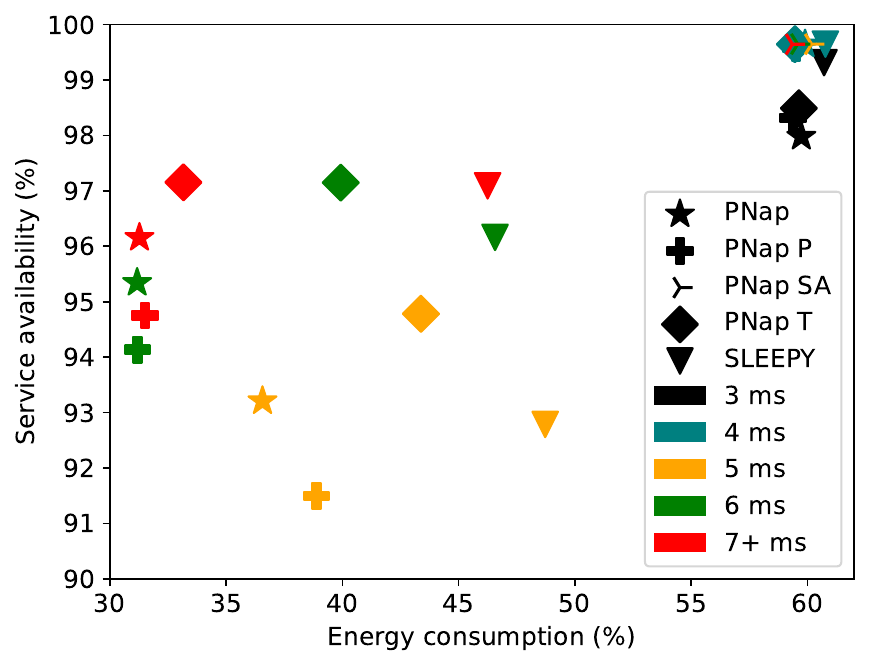}
  \caption{Trade-off between the ratio of energy consumed over peak energy consumption against  user-perceived service availability under increasing latency allowed.}
  \label{fig:pareto}
\end{figure}

Figure~\ref{fig:energy}  reports the ratio of energy consumed by \glspl{ec} with respect to the possible peak consumption; Figure~\ref{fig:sa} shows the service availability experienced by users. As expected, the Ideal approach achieves perfect service availability at minimal energy consumption, serving as an upper bound for performance. The Reactive approach, by contrast, highlights the drawbacks of neglecting \gls{ec} state transition delays: although its energy consumption appears comparable to the Ideal case, this comes at the cost of severe service disruptions,
making it effectively unable to serve the majority of requests at even moderate latency requirements. 

SLEEPY maintains high service availability within realistic lifecycle and energy management assumptions,  modestly reducing  energy consumption by  16.5\% as the latency budget increases. Its limited energy savings, however, reflect its inability to exploit deeper sleep states or anticipate future network dynamics.

In contrast, PNap exhibits a more balanced trade-off between service availability and energy consumption. Across all considered latency regimes, PNap's service availability is comparable to that of SLEEPY. This behavior can be attributed to the correct management of lifecycle oeprations. At the same time, PNap reduces energy consumption by 28.4\,\% when comparing the strictest and most relaxed latency constraints, which is 14.9\,\% more than the corresponding reduction observed for SLEEPY. These results indicate that the multi-state sleep and lifecycle-aware decision process of PNap makes performance more stable and reduces energy consumption without significant drawbacks with respect to SLEEPY.

\autoref{fig:state} presents how PNap approaches multi-state sleep, showing the average number of \glspl{ec} in a particular state as a function of the latency limit of the approach. Under strict latency constraints, most \glspl{ec} remain active to guarantee fewer forwarding hops. As soon as limited traffic rerouting becomes feasible without jeopardizing service quality, PNap progressively transitions more \glspl{ec} to deeper sleep states. When the latency limit reaches 5--6 ms, an increasing number of \glspl{ec} can safely enter the deepest sleep state, yielding significant energy savings. Beyond 7 ms, when forwarding distance no longer constitutes a limiting factor in this scenario, PNap almost closes the gap with the Ideal solver. In particular, in this scenario, PNap keeps active on average 1--2 \glspl{ec} more than the Ideal approach. This is explained by the realistic state transition delays, resulting in \glspl{ec} kept active to serve users while others prepare the required services. 

\autoref{fig:transition} shows the distribution the number of state transitions performed by \glspl{ec} under the PNap approach as the allowed latency increases.
The higher number of transitions when latency is limited to 5--6 ms explains the slight decrease in service availability observed in \autoref{fig:sa}, but at the same time enables PNap to achieve its substantial energy savings. As the allowed latency exceeds 7 ms and forwarding distance becomes less relevant, the average number of transitions drops to around one. In this case, most \glspl{ec} are placed in the S4 sleep state, with only a few nodes actively switching states.

Finally, to illustrate the adaptability of PNap to different operational objectives, \autoref{fig:pareto} reports the trade-off between energy consumption and service availability for four PNap variants as follows: 

\begin{itemize}
\item \textit{PNap}: the configuration considered so far; it focuses on minimizing energy consumption by reducing the number of active \glspl{ec}.
\item \textit{PNap P}: same policy as PNap, but the request processing priority is given to user requests instead of lifecycle requests.
\item \textit{PNap SA}: a configuration maximizing service availability. Traffic offloading is disabled and, as such, \glspl{ec} are allowed to enter sleep mode only when they would receive no traffic for prolonged periods.
\item \textit{PNap T}: a conservative configuration in which traffic offloading is limited, aiming to achieve higher service availability thanks to more active at the cost of higher energy consumption.
\end{itemize}

In particular, the two axis of the figure are the two metrics considered, different markers represent how different approaches score in those metrics and each color shows the latency allowed. Ideally, a good policy should have datapoints as close to the upper-left corner as possible, though this is not trivial due to the trade-off between energy consumption and service availability. 

As shown, PNap SA consistently achieves the highest service availability across all considered latency constraints, at the expense of higher and more stable energy consumption. Conversely, the baseline PNap configuration highlights the potential energy savings enabled by aggressive consolidation, while also exhibiting a noticeable reduction in service availability. Importantly, PNap P shows how important careful consideration of lifecycle is: if service state transitions have lower priority than user traffic services may not be ready when users need them, decreasing   service availability overall. 
PNap T lies between these two extremes, demonstrating that the operating point can be tuned to improve service availability compared to PNap, at the cost of an overall increase in energy consumption. In general moving the objective of the approach can be done by adjusting Algorithm~\autoref{algo2}, more specifically by modifying the offloading conditions. While not demonstrated in this study these adjustments can be, in principle, addressed at runtime to fit the current network conditions and strike a balance between the two examined metrics.

\section{Conclusions \& Outlook}
In this paper, we presented PNap, a \gls{stgcn}-based heuristic  for lifecycle-aware orchestration in \gls{mec} networks that jointly addresses multi-state sleep management and service lifecycle operations at \glspl{ec}. To the best of our knowledge, PNap is the first solution to explicitly consider these two aspects in \gls{mec} orchestration problems. By exploiting forecasts of future user connectivity over a finite time horizon, PNap identifies a reduced set of \glspl{ec} capable of sustaining user traffic and determines their most suitable activity or sleep state. This proactive decision-making enables timely service lifecycle operations, ensuring service continuity while reducing network energy consumption.

Performance evaluation results show that PNap achieves a favorable trade-off between energy efficiency and service availability. Specifically, PNap can reduce energy consumption by up to 28.4\,\% as the tolerated latency increases, while providing a 14.9\,\% improvement over the considered state-of-the-art solution. At the same time, it progressively closes the gap toward the behavior of an ideal solver that assumes perfect foresight and neglects practical constraints. Importantly, PNap matches the service availability of the state-of-the-art solution considered, demonstrating that these energy gains do not come at the cost of further service disruptions, but solely through correct sleep and lifecycle management. By tuning its parameters, PNap goes beyond a single standalone heuristic and can instead be viewed as a family of heuristics. In particular, adjusting the offloading policy allows trading energy consumption for service availability, and vice versa, making PNap adaptable to different network requirements.

PNap assumes the presence of a centralized orchestrator that performs estimations and decides for the entire network. But as the number of users, \glspl{ec}, and services grows, the computational and signaling overhead of a centralized approach could become prohibitive. For this reason, future work will focus on developing a distributed version of PNap, capable of scaling efficiently while retaining its benefits.

\bibliographystyle{IEEEtran}
\bibliography{lib.bib}

\newpage

\end{document}

%% file: acronyms.tex
\newacronym{5g}{5G}{Fifth-Generation Wireless Systems}
\newacronym{ai}{AI}{artificial intelligence}
\newacronym{ann}{ANN}{Artificial Neural Network}
\newacronym{ask}{ASK}{amplitude-shift keying}
\newacronym[plural={AWGN}, \glsshortpluralkey={AWGN}]{awgn}{AWGN}{additive white Gaussian noise}
\newacronym{au}{AU}{aerial user}
\newacronym{ap}{AP}{access point}
\newacronym{bec}{BEC}{binary erasure channel}
\newacronym{ber}{BER}{bit error rate}
\newacronym{bler}{BLER}{block error rate}
\newacronym{bpsk}{BPSK}{binary phase-shift keying}
\newacronym{bs}{BS}{base station}
\newacronym{bbu}{BBU}{baseband unit}

\newacronym{cdf}{CDF}{cumulative distribution function}
\newacronym{cnn}{CNN}{convolutional neural network}
\newacronym{csi}{CSI}{channel state information}
\newacronym{comp}{CoMP}{coordinated multi-point}
\newacronym{cran}{C-RAN}{cloud-radio access network}
\newacronym{cca}{CCA}{cloud coverage area}

\newacronym{dnn}{DNN}{deep neural network}
\newacronym{drl}{DRL}{deep reinforcement learning}
\newacronym{dqn}{DQN}{deep Q-network}
\newacronym{daps}{DAPS}{dual active protocol stack}
\newacronym{ecc}{ECC}{error-correcting code}
\newacronym{ecdf}{ECDF}{empirical distribution function}
\newacronym{ee}{EE}{energy efficiency}
\newacronym{ec}{EC}{edge cloud}
\newacronym{cc}{CC}{central cloud}
\newacronym{e2e}{E2E}{end-to-end}

\newacronym{ff}{FF}{feed-forward}
\newacronym{ffnn}{FF-NN}{feed-forward neural network}
\newacronym{fsk}{FSK}{frequency-shift keying}
\newacronym{fs}{FS}{functional split}
\newacronym{fsm}{FSM}{finite state machine}

\newacronym{gm}{GM}{Gaussian mixture}
\newacronym{GOPS}{GOPS}{Giga Operation Per Second}
\newacronym{gnn}{GNN}{Graph Neural Network}

\newacronym{hmappo}{H-MAPPO}{hierarchical multi-agent proximal policy optimization}
\newacronym{hdrl}{HDRL}{Hierarchical deep reinforcement learning}
\newacronym{hmarl}{HMARL}{Hierarchical Multi-Agent Reinforcement Learning}

\newacronym{iid}{i.i.d.}{independent and identically distributed}
\newacronym{il}{IL}{information leakage}
\newacronym{iot}{IoT}{internet of things}
\newacronym{ilp}{ILP}{Integer Linear Programming}

\newacronym{lb}{LB}{lower bound}
\newacronym{los}{LoS}{line-of-sight}
\newacronym{lstm}{LSTM}{long short-term memory}

\newacronym{mac}{MAC}{multiple access channel}
\newacronym{map}{MAP}{maximum a posteriori}
\newacronym{mc}{MC}{Monte Carlo}
\newacronym{mdp}{MDP}{Markov decision process}
\newacronym{mimo}{MIMO}{multiple-input multiple-output}
\newacronym{miso}{MISO}{multiple-input single-output}
\newacronym{ml}{ML}{machine learning}
\newacronym{mmwave}{mmWave}{millimeter wave}
\newacronym{mop}{MOP}{multi-objective programming problem}
\newacronym{mrc}{MRC}{maximum ratio combining}
\newacronym{msc}{MSC}{modulation and coding scheme}
\newacronym{mse}{MSE}{mean squared error}
\newacronym{msac}{MSAC}{masked soft actor-critic}
\newacronym{mappo}{MAPPO}{multi-agent proximal policy optimization}
\newacronym{madrl}{MADRL}{multi-agent deep reinforment learning}
\newacronym{milp}{MILP}{Mixed Integer Linear Programming}
\newacronym{mec}{MEC}{Multi-access Edge Computing}

\newacronym{nlos}{NLoS}{non-line-of-sight}
\newacronym{nn}{NN}{neural network}
\newacronym{nve}{NVE}{normalized validation error}
\newacronym{ric}{Near-RT RIC}{Near-Real Time RAN Intelligent Controller}
\newacronym{noma}{NOMA}{Non-orthogonal multiple access}
\newacronym{nfv}{NFV}{network function virtualization}
\newacronym{nlp}{NLP}{nonlinear programming}


\newacronym{oru}{O-RU}{O-RAN radio unit}
\newacronym{oran}{O-RAN}{open-radio access network}
\newacronym{odu}{O-DU}{O-RAN distributed unit}

\newacronym{pdf}{PDF}{probability density function}
\newacronym{pwc}{PWC}{polar wiretap code}
\newacronym{ppo}{PPO}{proximal policy optimization}
\newacronym{pf}{PF}{processing function}
\newacronym{qam}{QAM}{quadrature amplitude modulation}
\newacronym{qos}{QoS}{quality of service}

\newacronym{ra}{RA}{rearrangement algorithm}
\newacronym{rbf}{RBF}{radial basis function}
\newacronym{relu}{ReLU}{rectified linear unit}
\newacronym{ris}{RIS}{reconfigurable intelligent surface}
\newacronym{rl}{RL}{reinforcement learning}
\newacronym{rnn}{RNN}{recurrent neural network}
\newacronym{rf}{RF}{random forest}

\newacronym{sac}{SAC}{soft actor-critic}
\newacronym{sgd}{SGD}{stochastic gradient descent}
\newacronym{sgp}{SGP}{secure goodput}
\newacronym{simo}{SIMO}{single-input multiple-output}
\newacronym{sinr}{SINR}{signal-to-interference-plus-noise ratio}
\newacronym{siso}{SISO}{single-input single-output}
\newacronym{snr}{SNR}{signal-to-noise ratio}
\newacronym{svm}{SVM}{support vector machine}
\newacronym{sfc}{SFC}{service function chain}
\newacronym{sla}{SLA}{service level agreement}
\newacronym{stgcn}{STGCN}{spatio-temporal graph convolutional network}

\newacronym{tvd}{TVD}{total variation distance}
\newacronym{ttr}{TTR}{time to trigger}
\newacronym{tldqbd}{TLDQBD}{transient level-dependent quasi-birth-death process}

\newacronym{uav}{UAV}{unmanned aerial vehicle}
\newacronym{ub}{UB}{upper bound}
\newacronym{urllc}{URLLC}{ultra-reliable low latency communication}

\newacronym{v2v}{V2V}{vehicle-to-vehicle}
\newacronym{v2x}{V2X}{vehicle-to-everything}
\newacronym{vm}{VM}{virtual machine}
\newacronym{vnf}{VNF}{virtual network function}




\newacronym{zoc}{ZOC}{zero-outage capacity}